\documentclass[twocolumn]{svjour3}         
\smartqed  
\usepackage{graphicx}
 \usepackage{mathptmx}      
\usepackage{multirow}
\usepackage{amsmath}
\usepackage{amssymb}
\usepackage[round]{natbib}
 \journalname{}
\begin{document}

\title{Global sensitivity analysis of stochastic computer models
  with joint metamodels}

\titlerunning{Sensitivity analysis of stochastic models}        

\author{Bertrand Iooss         \and
        Mathieu Ribatet        \and
        Amandine Marrel
}


\institute{B. Iooss \at
              CEA, DEN, Centre de Cadarache, F-13108 Saint Paul lez Durance, France \\
              Tel.: +33-4-42257273\\
              Fax: +33-4-42252408\\
              \email{bertrand.iooss@cea.fr}           
           \and
           M. Ribatet \at
              Ecole Polytechnique F\'ed\'erale de Lausanne, Institute of
  Mathematics, Lausanne, Switzerland
           \and
           A. Marrel \at
           CEA, DEN, F-13108 Saint Paul lez Durance, France \\
             \emph{Present address:} Institut Fran\c{c}ais du P\'etrole, 92852 Rueil-Malmaison, France  
}

\date{Received: date / Accepted: date}

\maketitle

\begin{abstract}
The global sensitivity analysis method, used to quantify the influence of
  uncertain input variables on the response variability of a
  numerical model, is applicable to deterministic computer code (for
  which the same set of input variables gives always the same output
  value).  This paper proposes a global sensitivity analysis
  methodology for stochastic computer code (having a variability induced by
  some uncontrollable variables).  The framework of the
  joint modeling of the mean and dispersion of heteroscedastic data is
  used. To deal with the complexity of computer experiment outputs,
  non parametric joint models (based on Generalized
  Additive Models and Gaussian processes) are discussed. 
  The relevance of these new models is analyzed in terms of the obtained variance-based sensitivity indices with two case studies.
  Results show that the joint modeling approach leads accurate
  sensitivity index estimations even when clear heteroscedasticity is
  present.
\end{abstract}


\section{Introduction}

Many phenomena are modeled by mathematical equations which are
implemented and solved by complex computer code.  These computer
models often take as inputs a high number of numerical variables and
physical variables, and give several outputs (scalars or functions).
For the development of such computer models, its analysis, or its use,
the global Sensitivity Analysis (SA) method is an invaluable tool
\citep{salcha00,kle97,heljoh06}.  It takes into account all the variation ranges
of the inputs, and tries to apportion the output uncertainty to the
uncertainty in the input factors.  These techniques, often based on
the probabilistic framework and Monte-Carlo methods, require a lot of
simulations.  The uncertain input variables are modeled by random
variables and characterized by their probabilistic density functions.
The SA methods are used for model calibration \citep{kenoha01}, model validation \citep{bayber07}, decision
making process \citep{derdev08}, i.e. all the processes where it is
useful to know which variables mostly contribute to output
variability . 

The current SA methods are applicable to the deterministic computer
code, i.e. for which the same set of input variables always gives the
same output values.  The randomness is limited to the model inputs,
whereas the model itself is deterministic.  
For example in the nuclear engineering domain,
global sensitivity analysis tools have been applied to waste storage
safety studies \citep{heljoh06},
and pollutant transport modeling in the
aquifer \citep{volioo08}. In such
industrial studies, numerical models are often too time consuming for 
applying directly the global SA methods.  To avoid this problem, one
solution consists in replacing the time consuming computer code by an
approximate mathematical model, called metamodel
\citep{sacwel89,fanli06}.  This
function must be as representative as possible of the computer code,
with good prediction capabilities and must require a negligible
calculation time.  Several metamodels are classically used:
polynomials, splines, neural networks, Gaussian processes
\citep{chetsu06,fanli06}.

In this paper, we are not concerned by deterministic computer models
but by stochastic numerical models - i.e. when the same input
variables set leads to different output values.  The model itself
relies on probabilistic methods (e.g. Monte-Carlo) and is
therefore intrinsically stochastic because of some ``uncontrollable variables''.
For the uncertainty analysis, \citet{kle97} has
raised this question, giving an example concerning a queueing
model. In the nuclear engineering domain, examples are given by
Monte-Carlo neutronic models used to calculate elementary particles
trajectories, Lagrangian stochastic models for simulating a large
number of particles inside turbulent media (in atmospheric or
hydraulic environment).  In our study, ``uncontrollable'' variables
correspond to variables that are known to exist, but unobservable,
inaccessible or non describable for some reasons.  It includes the
important case in which observable vector variables are too
complex to be described or synthesized by a reasonable number of scalar parameters.
This last situation might concern the code for which some simulations
of complex random processes are used.  For example, one can quote some
partial differential equation resolutions in heterogeneous random
media simulated by geostatistical techniques (\citealp[e.g. fluid flows in oil
reservoirs,][]{zabdej98}, \citealp[and acoustical wave
propagation in turbulent fluids,][]{ioolhu02}), where
the uncontrollable variable is the simulated spatial field involving
several thousand scalar values for each realization.
Of course, in this case, behind this uncontrollable variable stands a fully
controllable parameter: the random seed. However, the effect of the random seed
on the computer code output is totally chaotic because a slight modification of the random seed
leads to a very different random medium realization.
For simplicity and for generality, we use the expression ``uncontrollable variable'' in this paper.


For stochastic computer models, classical metamodels (devoted to
approximate deterministic computer models) are not pertinent anymore.
To overcome this problem, the commonly used Gaussian process (Gp)
model \citep{sacwel89,marioo07} can
include an additive error component (called the ``nugget effect'') by
adding a constant term into its covariance function.  However, it supposes that the error term
is independent of the input variables (homoscedasticity hypothesis),
which means that the uncontrollable variable does not interact with
controllable variables.  This hypothesis limits the use of Gp to
specific cases even if recently, some authors
\citep[e.g.][]{klevan05} demonstrated the usefulness of Gp for stochastic
computer model in heteroscedastic cases.  

To construct heteroscedastic
metamodels for stochastic computer code, \citet{zabdej98} have proposed another approach by modeling the
mean and the dispersion (i.e. the variance) of computer code outputs
by two interlinked Generalized Linear Models (GLMs).  This approach,
called the joint model, has been previously studied in the context of
experimental data modeling \citep{smy89,macnel89}.

Modeling the mean and variance of a response variable in function of
some controllable variables is of primary concern in product
development and quality engineering methods \citep{pha89}:
experimentation is used to determine the factor levels so that the
product is insensitive to potential variations of environmental
conditions.  This can be summarized, in the framework of the robust
design, as the optimization of a mean response function while
minimizing a variance function.  In this context, \citet{vinmye90} propose to build polynomial models for the mean and
the variance separately, while \citet{leenel03} consider
the joint GLM approach.  A recent and complete review on this subject
can be found in \citet{burste06}.  Dealing with
computer experiments instead of physical ones, \citet{batken06} propose different strategies for designing and
analyzing robust design experiments. In this case, the noise factors
are fully controllable. This allows the authors to provide a powerful
stochastic emulator strategy.

Following the work of \citet{zabdej98}, \citet{ioorib06,ioorib08} have recently
introduced the joint model to perform a global sensitivity analysis of
a stochastic computer code. Results show that a total sensitivity index of all
the uncontrollable variables can be computed using the dispersion
component of the joint model.  However, the parametric form of the GLM
framework provides some limitations when modeling complex computer
code outputs. To bypass this hurdle, this paper suggests the use of
non parametric models.  Due to its similarity with GLMs, Generalized
Additive Models (GAM) are considered \citep{hastib90,wooaug02}, even though Gp or
other non parametric models should also be a relevant solution.

This paper starts by describing the joint model construction, firstly
with the GLM, secondly with the GAM. We will also show how other
models, like Gp, can be used to model the mean and dispersion
components.  The third section describes the global sensitivity
analysis for deterministic models, and its extension to stochastic
models using joint models.  Particular attention is devoted to the
calculation of variance-based sensitivity indices (the so-called Sobol
indices).  Considering a simple analytic function, the performance of
the proposed approach is compared to other commonly used models.
Next, an application on an actual industrial case (groundwater
radionuclide migration modeling) is given.  Finally, some conclusions
synthesize the contributions of this work.


\section{Joint modeling of mean and dispersion}\label{par2} 

\subsection{Using the Generalized Linear Models}


The class of GLM allows to extend the class of the traditional linear
models by the use of: (a) a distribution which belongs to the
exponential family; (b) and a link function which connects the
explanatory variables to the explained variable \citep{nelwed72}.  Let us describe the first component of the model
concerning the mean:
\begin{equation}\label{eqGlmM}
  \begin{cases}
    \mathbb{E}(Y_i) & = \mu_i,\qquad \eta_i = g(\mu_i) = \sum_j
    x_{ij}\beta_j \;,\\ 
    \mbox{Var}(Y_i) &= \phi_i v(\mu_i) \;,
  \end{cases}
\end{equation}
where $(Y_i)_{i=1\ldots n}$ are independent random variables with mean
$\mu_i$; $x_{ij}$ are the observations of the variable $X_j$;
$\beta_j$ are the regression parameters which have to be estimated;
$\eta_i$ is the mean linear predictor; $g(\cdot)$ is a differentiable
monotonous function (called the link function); $\phi_i$ is the
dispersion parameter and $v(\cdot)$ is the variance function.  To
estimate the mean component, the functions $g(\cdot)$ and $v(\cdot)$
have to be specified.  Some examples of link functions are given by
the identity (traditional linear model), root square, logarithm, and
inverse functions. Some examples of variance functions are given by
the constant (traditional linear model), identity and square
functions.

Within the joint model framework, the dispersion parameter $\phi_i$ is
not supposed to be constant as in a traditional GLM,
but is supposed to vary according to the model:
\begin{equation}\label{eqGlmD}
  \begin{cases}
    \mathbb{E}(d_i) &= \phi_i,\qquad \zeta_i = h(\phi_i) = \sum_j
    u_{ij} \gamma_j \;,\\ 
    \mbox{Var}(d_i) &= \tau v_d(\phi_i) \;,
  \end{cases}
\end{equation}
where $d_i$ is a statistic representative of the dispersion,
$\gamma_j$ are the regression parameters which have to be estimated,
$h(\cdot)$ is the dispersion link function, $\zeta_i$ is the
dispersion linear predictor, $\tau$ is a constant and $v_d(\cdot)$ is
the dispersion variance function.  $u_{ij}$ are the observations of
the explanatory variable $U_j$.  The variables $(U_j)$ are generally
taken among the explanatory variables of the mean $(X_j)$, but might
differ.  To ensure positivity, $h(\phi)=\log \phi$ is often taken for
the dispersion link function while the statistic $d$ representing the
dispersion is generally taken to be the deviance contribution - which
is approximately $\chi^2$ distributed.  Therefore, as the $\chi^2$
distribution is a particular case of the Gamma distribution,
$v_d(\phi)=\phi^2$ and $\tau \sim 2$.
  
The joint model is fitted using the Extended Quasi Loglikelihood
\citep[EQL,][]{nelpre87} maximization. The EQL behaves as
a log-likelihood for both mean and dispersion parameters.
This justifies an iterative procedure to fit the joint model. First, a
GLM is fitted on the mean; then from the estimate of $d$, another GLM
is fitted on the dispersion.  From the estimate of $\phi$, weights for
the next estimate of the GLM on the mean are obtained.  This process
can be reiterated as many times it is necessary, and allows to
entirely fit our joint model \citep{macnel89}.

Statistical tools available in the GLM fitting are also available for
each component of the joint model: deviance analysis and Student test.
It allows to make some variable selection in order to simplify model
expressions.  The residuals graphical analysis (which have to be
normally distributed) and the q-q plots can be used as indicators of
the correctness of the link function for the mean component
\citep{leenel03}.  In practice, some evidence can lead to an
adequate choice of the link function \citep{macnel89}.  For example, a binomial-type explained variable
leads to the use of the logit function.  However, if a natural choice
is not possible and if the identity link function does not provide
satisfactory residuals analysis, plotting the adjusted dependent
variable versus the linear predictor might help in choosing a more
appropriate link function \citep{macnel89}.

In conclusion, all the results obtained on the joint GLM are applicable to
the problem of stochastic computer experiments.
The novelty proposed in our paper concerns the global sensitivity analysis
issue (section \ref{sec:gsasm}). Moreover, in the following section we
extend the joint GLM to the non parametric framework.
This kind of model is necessary for the computer experiment outputs
which tend to be rather complex and need non parametric modeling.

{\it 
\noindent{\bf \itshape Remark:}
A simpler approach consists in building polynomial models for the mean
and the variance separately \citep{vinmye90,burste06}.  This approach, called dual modeling,
consists in repeating calculations with the same sets of controllable
variables (which is not necessary in the joint modeling approach).
The dual modeling approach has been successfully applied in many
situations, especially for robust conception problems.  However for
our purpose (accurate fitting of the mean and dispersion components),
it has been shown that this dual model is less competitive than the
joint model \citep{zabdej98,leenel03}:
  the dual modeling approach fits the dispersion model
  given the mean model and this approach does not always lead to
  optimal fits. 
  }

\subsection{Extension to the Generalized Additive Models}

Generalized Additive models (GAM) were introduced by 
\citet{hastib90} and allow a linear term in the
linear predictor $\eta=\sum_j \beta_j X_j$ of equation (\ref{eqGlmM})
to be replaced by a sum of smooth functions $\eta=\sum_j s_j(X_j)$.
The $s_j(.)$'s are unspecified functions that are obtained by fitting
a smoother to the data, in an iterative procedure.  GAMs provide a
flexible method for identifying nonlinear covariate effects in
exponential family models and other likelihood-based regression
models.  The fitting of GAM introduces an extra level of iteration in
which each spline is fitted in turn assuming the others known.  GAM
terms can be mixed quite generally with GLM terms in deriving a model.

One common choice for $s_j$ is the smoothing spline, i.e. splines with
knots at each distinct value of the variables.  In regression
problems, smoothing splines have to be penalized in order to avoid
data overfitting. \citet{wooaug02} have described in
details how GAMs can be constructed using penalized regression
splines.  Because numerical models often exhibit strong interactions
between input variables, the incorporation of multi-dimensional
smooth (for example the bi-dimensional spline term $s_{ij}(X_i,X_j)$)
is particularly important in our context.

GAMs are generally fitted using penalized likelihood maximization. For
this purpose, the likelihood is modified by the addition of a penalty
for each smooth function, penalizing its ``wiggliness''. Namely, the
penalized loglikelihood is defined as:
\begin{equation}
  \label{eq:pllik}
  p\ell = \ell + \sum_{j=1}^p \lambda_j \int \left(\frac{\partial^2
      s_j}{\partial x_j^2} \right)^2 dx_j
\end{equation}
where $\ell$ is the loglikelihood function, $p$ is the total number of
smooth terms and $\lambda_j$ are ``smoothing parameters'' which
compromise between goodness of fit and smoothness.

Estimation of these ``smoothing parameters'' is generally achieved using
the GCV score minimization. The GCV score is defined as:
\begin{equation}
  \label{eq:gcvScore}
  S_{GCV} = \frac{n d}{\left(n - DoF\right)^2}
\end{equation}
where $n$ is the number of data, $d$ is the deviance and $DoF$ is the
effective degrees of freedom, i.e. the trace of the so-called ``hat''
matrix. Extension to (E)QL models is straightforward by substituting
the likelihood function and the deviance $d$ for their (extended)
quasi counterparts. In practice, all the smoothing parameters are jointly updated
at each iteration of the fitting procedure of the joint model. To this
aim, on every iteration a GLM/GAM is fitted for each trial set of
smoothing parameters, and GCV scores are only evaluated at
convergence.

We have seen that GAMs extend in a natural way GLMs.  Therefore, it
would be interesting to extend the joint GLM model to a joint GAM
one. Such ideas have been proposed in 
\citet{rigsta96} where both the mean and variance were modeled using
semi-parametric additive models 
\citep{hastib90}.  This model is restricted to observations following
a Gaussian distribution and is called Mean and Dispersion Additive
Model (MADAM). Our model is more general and relaxes the Gaussian
assumption as now quasi-distributions are considered: while
the MADAM fitting procedure relies on the maximization of the
penalized likelihood, the joint GAM maximizes the penalized extended
quasi-likelihood. In addition, \citet{rigsta96}
only used cubic regression splines, while our framework allows also the use
of multivariate smoothers - e.g. thin plate regression splines.
As our model is based on GAMs and by analogy with the denomination
``joint GLM'', we call it ``joint GAM'' in the following.


Lastly, it has to be noticed that, within the EQL maximization
framework, a large number of models can be considered instead of
GAMs. For instance, one can use a GAM for the mean response and a GLM
for the dispersion component. In addition, more complex models can
also be considered such as Gaussian processes - see Section
\ref{secother}.

\subsection{Joint modeling with other models}\label{secother}

For some applications, joint GAM could be inadequate, and other models
can be proposed.  For example, for Gaussian observations, 
\citet{juuron05} have used a neural network model for mean
and dispersion.  It is shown to be more efficient than joint GLM and
joint additive models in a context of numerous explanatory variables
($25$) and of a large amount of data ($100000$).  They perform an
extensive comparison for large data sets between joint neural network
model, MADAM, joint local linear regression model and joint linear
regression model.  While our context of computer experiments is
different (we have small data sets), it is interesting to recall their
conclusion:
\begin{itemize}
\item the neural network joint model gives the best prediction performance;
\item MADAM requires a huge amount of memory;
\item joint local linear model is extremely time consuming;
\item joint linear model is appropriate when simplicity is required.
\end{itemize}

It is also possible to build a heteroscedastic model based on the
Gaussian process (Gp) metamodel \citep[also known as the kriging principle,][]{sacwel89}.
The Gp approach essentially is a kind of linear interpolation built on the property
of the multivariate normal distribution.
Gp metamodel gives
not only a predictor (which is the best linear unbiased predictor) of a computer experiment but also a local
indicator of prediction accuracy.
For heteroscedastic data, a
first approach, proposed by \citet{ginrou08},
consists in modeling the mean of the computer code with a Gp metamodel
for which the nugget effect is supposed to vary with the inputs. 
From this fitted Gp, one can use the estimation of the MSE (given by the Gp model)
as the  dispersion statistic $d$ introduced in Equation (\ref{eqGlmD}).
  This model does not require
any fitting of the dispersion component and we prefer to focus our
attention on another method, the joint Gp model, which is coherent
with our previous joint models.
\citet{boucor09} have 
recently introduced a such joint Gp model for the same purpose.

The first step of our methodology models the mean by a Gp metamodel
(having a nugget effect) estimated on the learning sample.  The second
step consists in adjusting a second Gp metamodel on the squared
residuals.  This process can be iterated as in the joint GLM and joint
GAM fitting procedure.  Due to the presence of a nugget effect in the
mean component, the mean Gp is not anymore an exact interpolator and
the learning sample residuals can be used for the dispersion model.
However, residuals could also be derived from a cross validation
method.


\section{Global sensitivity analysis}

\subsection{Deterministic models}

The global SA methods are applicable to deterministic computer code,
e.g. for which the same set of input variables always leads to the
same response value.  This is considered by the following model:
\begin{equation}\label{eqmodeldeterm}
  \begin{array}{rcl}
    f : & \mathbb{R}^p & \rightarrow \mathbb{R} \\
    & \mathbf{X} & \mapsto Y=f(\mathbf{X})
  \end{array}
\end{equation}
where $f(\cdot)$ is the model function (possibly analytically unknown), 
$\mathbf{X}=(X_1,\ldots,X_p)$ are $p$
independent inputs and $Y$ is the output.
In our problem, $\mathbf{X}$ is uncertain and considered as a 
random vector with known distribution which reflects this uncertainty.
Therefore, $Y$ is also a random variable, whose distribution is unknown. 
In this section, let us recall some basic
ideas on the variance-based sensitivity indices, called Sobol indices,
 applied on this model.

Among quantitative methods, variance-based methods are the most often
used \citep{salcha00}.  The main idea of these
methods is to evaluate how the variance of an input or a group of
inputs contributes into the variance of output.  We start from the
following variance decomposition:
\begin{eqnarray}\label{vartotale} 
  \mbox{Var} \left[Y \right]= \mbox{Var} \left[\mathbb{E}\left(Y|X_i
    \right) \right] + \mathbb{E} \left[\mbox{Var} \left(Y|X_i\right)
  \right] \;,
\end{eqnarray} 
which is known as the total variance theorem.  The first term of this
equality, named variance of the conditional expectation, is a natural
indicator of the importance of $X_i$ into the variance of $Y$: the
greater the importance of $X_i$, the greater is
$\mbox{Var}[\mathbb{E}(Y|X_i)]$. Most often, this term is divided by
$\mbox{Var}[Y]$ to obtain a sensitivity index in $[0,1]$.

To express the sensitivity indices, we use
the unique decomposition of any integrable function on $[0,1]^p$
 into a sum of elementary functions \citep[see for example][]{sob93}:
 \begin{equation}
\begin{array}{r}
  \label{decompsobol}
 \displaystyle f(X_1,\cdots,X_p) = f_0 + \sum_i^p f_i(X_i) + \sum_{i<j}^p f_{ij}(X_i,X_j) \\
 \displaystyle+ \ldots +   f_{12..p}(X_1,\cdots,X_p) \;,
\end{array}
\end{equation}
where $f_0$ is a constant and the other functions verify the following conditions:
\begin{eqnarray} 
  \int_0^1 f_{i_1,\ldots,i_s}(x_{i_1},\ldots,x_{i_s})\,dx_{i_k}=0 
\end{eqnarray}
$\forall k=1,\ldots,s\;,\;\; \forall \{i_1,\ldots,i_s\}\subseteq \{1,\ldots,p\}$.
If the $X_i$s are mutually independent, the decomposition (\ref{decompsobol}) is valid for any distribution functions for the $X_i$s.

From (\ref{decompsobol}), the following decomposition of the model output
variance is possible \citep{sob93}:
\begin{equation}
\begin{array}{r}
  \label{decompvar}
 \displaystyle \mbox{Var}\left[Y\right] = \sum_i^p V_i(Y) + \sum_{i<j}^pV_{ij}(Y)  + \sum_{i<j<k}^p V_{ijk}(Y) \\
 \displaystyle + \ldots +V_{12..p}(Y) \;,
\end{array}
\end{equation}
where $V_i(Y)=\mbox{Var}[\mathbb{E}(Y|X_i)]$,
$V_{ij}(Y)=\mbox{Var}[\mathbb{E}(Y|X_iX_j)]-V_i(Y)-V_j(Y),  \ldots\;$
 One
can thus define the sensitivity indices by:
\begin{equation}
\begin{array}{l}
  \label{eqindordre1}
\displaystyle  S_i=\frac{\mbox{Var}\left[\mathbb{E}\left(Y|X_i\right)\right]}{\mbox{Var}(Y)}
  =  \frac{V_i(Y)}{\mbox{Var}(Y)}, \\
\displaystyle  \quad  S_{ij} =
  \frac{V_{ij}(Y)}{\mbox{Var}(Y)}, \quad  S_{ijk} =
  \frac{V_{ijk}(Y)}{\mbox{Var}(Y)}, \quad \ldots  
\end{array} 
\end{equation}
These coefficients are called the Sobol indices, and can be used for
any complex model functions $f$.  
The second order index $S_{ij}$ expresses sensitivity
of the model to the interaction between the variables $X_i$ and $X_j$
(without the first order effects of $X_i$ and $X_j$), and so on for
higher orders effects.  The interpretation of these indices
is natural as their sum is equal to one (thanks to equation
(\ref{decompvar})): the larger and close to one an index value, the
greater is the importance of the variable or the group of variables
linked to this index.


For a model with $p$ inputs, the number of Sobol indices is $2^p-1$;
leading to an intractable number of indices as $p$ increases. Thus, to
express the overall sensitivity of the output to an input
$X_i$, \citet{homsal96} introduce the total
sensitivity index:
\begin{eqnarray}\label{indtotal}
  S_{T_i} = S_i + \sum_{j \neq i} S_{ij} + \sum_{j \neq i, k \neq i,
    j<k} S_{ijk} + \ldots = \sum_{l \,\in \,\# i}S_l \;, 
\end{eqnarray}
where $\# i$ represents all the ``non-ordered'' subsets of indices
containing index $i$.  Thus, $\sum_{l \,\in \,\# i}S_l$ is the sum of
all the sensitivity indices containing $i$ in their index.  For
example, for a model with three input variables,
$S_{T_1}=S_1+S_{12}+S_{13}+S_{123}$.

The estimation of these indices can be done by Monte-Carlo simulations
or by alternative methods \citep{salcha00}.  Recent algorithms have also been
introduced to reduce the number of required model evaluations
significantly.  As explained in the introduction, a powerful
method consists in replacing complex computer models by metamodels
which have negligible calculation time \citep[e.g.][]{volioo08}.  Estimation of Sobol indices
by Monte-Carlo techniques with their confidence intervals (requiring
thousand of simulations) can then be done using these response
surfaces. 

\subsection{Stochastic models}\label{sec:gsasm}

In this work, models containing some intrinsic alea, which is
described as an uncontrollable random input variable $\varepsilon$,
are called ``stochastic computer models''.  
Let us recall the example proposed in the introduction where
$\varepsilon$ is a random field whose each realization is governed by a random seed value.
We consider the random field $\varepsilon$ as an uncontrollable variable when this
random field is too
complex to be described or synthesized by a reasonable number of scalar parameters.

In the following, the expectation and variance operators involve averaging over the 
distribution of $(\mathbf{X},\varepsilon)$, unless another distribution is indicated.
Similarly from equation
(\ref{eqmodeldeterm}), consider the following (stochastic) model:
\begin{equation}\label{eqmodelstoch}
  \begin{array}{rcl}
    g : & \mathbb{R}^p & \rightarrow \mathbb{R} \\
    & \mathbf{X} & \mapsto Y=f(\mathbf{X}) +
    \nu(\varepsilon,\mathbf{X}:\varepsilon)  
  \end{array}
\end{equation}
where $\mathbf{X}$ are the $p$ controllable input variables
(independent random variables), $Y$ is the output, $f(\cdot)$ is the
deterministic part of the model function and
$\nu(\cdot)$ is the stochastic part of the model function.
Let $\mathbb{E}_\varepsilon(\nu)=0$ which means that $\nu(\cdot)$ is centered relatively to $\varepsilon$:  we put inside $f(\cdot)$ a possible constant term involved by $\nu(\cdot)$.  The notation
$\nu(\varepsilon,\mathbf{X}:\varepsilon)$ means that $\nu$ depends
only on $\varepsilon$ and on the interactions between $\varepsilon$
and $\mathbf{X}$.  The additive form of equation (\ref{eqmodelstoch})
is deduced directly from the decomposition of the function $g$ into
a sum of elementary functions depending on $(\mathbf{X},\varepsilon)$ 
(like the decomposition in Eq. (\ref{decompsobol})).

For a stochastic model (\ref{eqmodelstoch}),
the joint models introduced in section \ref{par2} enables us to recover
two GLMs, two GAMs or two Gps:
\begin{equation}\label{eqYm}
  Y_m(\mathbf{X}) = \mathbb{E}(Y|\mathbf{X})~=~ \mathbb{E}_\varepsilon(Y|\mathbf{X}) 
\end{equation}
by the mean component (Eq. (\ref{eqGlmM})), and
\begin{equation}\label{eqYd}
  Y_d(\mathbf{X}) = \mbox{Var}(Y|\mathbf{X}) = \mbox{Var}_\varepsilon(Y|\mathbf{X}) 
\end{equation}
by the dispersion component (Eq. (\ref{eqGlmD})).  If there is no
uncontrollable variable $\varepsilon$, it leads to a deterministic
model case with $Y_d(\mathbf{X})=\mbox{Var}(Y|\mathbf{X})=0$.  By using the total
variance theorem (Eq. (\ref{vartotale})), the variance of the output
variable $Y$ can be decomposed by:
\begin{equation}
\begin{array}{rcl}
  \label{decompvarcond}
  \mbox{Var}(Y) & = &
  \displaystyle \mbox{Var}_{\mathbf{X}}\left[\mathbb{E}\left(Y|\mathbf{X}\right) \right] +
  \mathbb{E}_{\mathbf{X}}\left[\mbox{Var}\left(Y|\mathbf{X}\right) \right] \\
  & = &
  \displaystyle \mbox{Var}_{\mathbf{X}}\left[Y_m(\mathbf{X}) \right] + \mathbb{E}_{\mathbf{X}}\left[Y_d(\mathbf{X}) \right]  \;. 
\end{array}
\end{equation}
According to model (\ref{eqmodelstoch}), $Y_m(\mathbf{X})$ is the deterministic
model part, and $Y_d(\mathbf{X})$ is the variance of the stochastic model part:
\begin{equation}
  \begin{array}{rcl}
    Y_m(\mathbf{X}) & = & f(\mathbf{X}) \;,\\
    Y_d(\mathbf{X}) & = & \mbox{Var}_\varepsilon \left[\nu(\varepsilon,\mathbf{X}:
    \varepsilon)|\mathbf{X} \right] \;.
  \end{array}
\end{equation}

The variances of $Y$ and $Y_m(\mathbf{X})$ are now decomposed according to the
contributions of their input variables $\mathbf{X}$.  For $Y$, the
same decomposition than for deterministic models holds (Eq.
(\ref{decompvar})).  However, it includes the additional term
$\mathbb{E}_{\mathbf{X}}\left[Y_d(\mathbf{X}) \right]$ (the mean of the dispersion component)
deduced from equation (\ref{decompvarcond}).  Consequently,
\begin{equation}
\begin{array}{r}
  \displaystyle \mbox{Var}(Y)=\sum_i^pV_i(Y) + \sum_{i<j}^pV_{ij}(Y) +
  \sum_{i<j<k}^p V_{ijk}(Y) \\
 \displaystyle + \ldots +V_{12..p}(Y) + \mathbb{E}_{\mathbf{X}}\left[Y_d(\mathbf{X}) \right]\;,
\end{array}
\end{equation}
where $V_i(Y)=\mbox{Var}_{X_i}[\mathbb{E}(Y|X_i)]$,
$V_{ij}(Y)=\mbox{Var}_{(X_i,X_j)}[\mathbb{E}(Y|X_iX_j)]-V_i(Y)-V_j(Y),  \ldots\;$
For the mean component $Y_m(\mathbf{X})$ that we note $Y_m$ for easing the notation, we have
\begin{equation}
\begin{array}{r}
\displaystyle  \mbox{Var}(Y_m)=\sum_i^pV_i(Y_m) + \sum_{i<j}^pV_{ij}(Y_m) +
  \sum_{i<j<k}^p V_{ijk}(Y_m) \\
  \displaystyle + \ldots +V_{12..p}(Y_m) \;. 
\end{array}
\end{equation}
By noticing that 
\begin{equation}
\begin{array}{rcl}
\displaystyle V_i(Y_m) & = & \displaystyle  \mbox{Var}_{X_i}[\mathbb{E}_{\mathbf{X}}(Y_m|X_i)
] = \mbox{Var}_{X_i}\{\mathbb{E}_\mathbf{X}[\mathbb{E}_\varepsilon(Y|\mathbf{X})|X_i]\} \\
& = & \displaystyle \mbox{Var}_{X_i}[\mathbb{E}_{\mathbf{X},\varepsilon}(Y|X_i) ] = V_i(Y)\;,
\end{array}
\end{equation}
 and from equation
(\ref{eqindordre1}), the sensitivity indices for the variable
$Y$ according to the controllable variables
$\mathbf{X}=(X_i)_{i=1 \ldots p}$ can be computed using:
\begin{equation}\label{eqSiY} S_i = \frac{V_i(Y_m)}{\mbox{Var}(Y)},
  \quad S_{ij}=\frac{V_{ij}(Y_m)}{\mbox{Var}(Y)}, \quad \ldots \;
\end{equation} 
These Sobol indices can be computed by classical Monte-Carlo
techniques, the same ones used in the deterministic model case.  These
algorithms are applied on the  metamodel defined by the mean
component $Y_m$ of the joint model.

Thus, all terms contained in $\mbox{Var}_\mathbf{X}[Y_m(\mathbf{X})]$
of the equation (\ref{decompvarcond}) have been considered.  It
remains to estimate $\mathbb{E}_\mathbf{X}[Y_d(\mathbf{X})]$ by a
simple numerical integration of $Y_d(\mathbf{X})$ following the
distribution
of $\mathbf{X}$.  $Y_d(\mathbf{X})$ is evaluated with a metamodel, for
example the dispersion component of the joint model.
$\mathbb{E}_\mathbf{X}[Y_m(\mathbf{X})]$ includes all the
decomposition terms of $\mbox{Var}(Y)$ (according to $\mathbf{X}$ and
$\varepsilon$) not taken into account in
$\mbox{Var}_\mathbf{X}[Y_m(\mathbf{X})]$ i.e. all terms involving
$\varepsilon$.  Therefore, the total sensitivity index of
$\varepsilon$ is
\begin{equation}\label{eqSTeps} 
  S_{T_\varepsilon} = \frac{\mathbb{E}_\mathbf{X}[Y_m(\mathbf{X})]}{\mbox{Var}(Y)} \;. 
\end{equation} 
As $Y_d(\mathbf{X})$ is a positive random variable, positivity of
$S_{T_\varepsilon}$ is guaranteed.
In practice, $\mbox{Var}(Y)$ can be estimated from the data or from
simulations of the fitted joint model:
\begin{equation}\label{eqvarY}
\mbox{Var}(Y)=\mbox{Var}_\mathbf{X}[Y_m(\mathbf{X})]+\mathbb{E}_\mathbf{X}[Y_m(\mathbf{X})] \;.
\end{equation}
If $\mbox{Var}(Y)$
is computed from the data, it seems preferable to estimate
$\mathbb{E}_\mathbf{X}[Y_m(\mathbf{X})]$ with $\mbox{Var}(Y)-\mbox{Var}_\mathbf{X}[Y_m(\mathbf{X})]$ to satisfy
equation (\ref{decompvarcond}).  In our applications, the total
variance will be estimated using the fitted joint model (Eq. (\ref{eqvarY})).

Finally, let us note that we cannot quantitatively
distinguish the various contributions in $S_{T_\varepsilon}$
($S_{\varepsilon}$, $S_{i\varepsilon}$, $S_{ij\varepsilon}$, \ldots).
Indeed, it is not possible to combine the functional anova decomposition
of $Y_m(\mathbf{X})$ with the functional anova decomposition of $Y_d(\mathbf{X})$ in
order to deduce the unknown sensitivity indices.
Finding a way to form some composite indices still remains
an open problem which needs further research.
However, we argue that the analysis of the terms in the regression model $Y_d$ and
their $t$-values give useful qualitative information.  For example, if
an input variable $X_i$ is not present in $Y_d$, we can deduce the
following correct information: $S_{i\varepsilon}=0$.
Moreover, if the $t$-values analysis and the deviance analysis show
that an input variable $X_i$ has a smaller influence than another
input variable $X_j$ , we can suppose that the interaction between
$X_i$ and $\varepsilon$ is less influential than the interaction between
$X_j$ and $\varepsilon$.  

In conclusion, this new approach, based on joint models to compute
Sobol sensitivity indices, is useful if the following conditions hold:
\begin{itemize}
\item if the computer model contains some uncontrollable variables
(the model is no more deterministic but stochastic);
\item if a metamodel is needed due to large CPU times of the computer model;
\item if some of the uncontrollable variables interact with some
  controllable input ones;
\item if some information about the influence of the interactions
  between the uncontrollable variables and the other input variables
  is of interest.
\end{itemize}



\section{Applications}

\subsection{An analytic test case: the Ishigami function}

The proposed method is first illustrated on an artificial analytical
model with $3$ input variables, called the Ishigami function \citep{homsal96,salcha00}:
\begin{equation}\label{eq:Ishigami}
  Y=f(X_1,X_2,X_3) = \sin(X_1)+7\sin(X_2)^2+0.1X_3^4\sin(X_1)
\end{equation}
where $X_i\sim \cal{U}[-\pi;\pi]$ for $i=1,2,3$.  For this function,
all the Sobol sensitivity indices ($S_1$, $S_2$, $S_3$, $S_{12}$,
$S_{13}$, $S_{23}$, $S_{123}$, $S_{T_1}$, $S_{T_2}$, $S_{T_3}$) are
known.  This function is used in most intercomparison studies of
global sensitivity analysis algorithms.  In our study, the classical
problem is altered by considering $X_1$ and $X_2$ as the controllable
input random variables, and $X_3$ as an uncontrollable input random
variable.  It means that the $X_3$ random values are not used in the
modeling procedure; this variable is considered to be inaccessible.
However, sensitivity indices have the same theoretical values as in
the standard case.

For this analytical function case, it is easy to obtain the exact mean and dispersion models
by deriving (via analytical integration) the analytical expressions of
the mean component $Y_m(X_1,X_2)$ and dispersion component $Y_d(X_1,X_2)$:
\begin{equation}\label{espvarishanal}
  \begin{array}{rcl}
Y_m(X_1,X_2) & = & \mathbb{E}(Y|X_1,X_2) \\
& = & \displaystyle \left(1+\frac{\pi^4}{50}\right)\sin(X_1)+7[\sin(X_2)]^2 \;, \\
Y_d(X_1,X_2) & = & \mbox{Var}(Y|X_1,X_2) \\
& = & \displaystyle \pi^8 \left(\frac{1}{900}-\frac{1}{2500} \right)[\sin(X_1)]^2 = Y_d(X_1) \;.
\end{array}
\end{equation}

\subsubsection{Metamodeling}\label{parmetamodeling}

For the model fitting, $1000$ Monte Carlo samples of $(X_1,X_2,X_3)$ were
simulated leading to $1000$ observations for $Y$.
No replication is made in the $(X_1,X_2)$ plane
because it has been shown that repeating calculations with the same sets of controllable
variables is inefficient in the joint modeling approach \citep{zabdej98,leenel03}.
Therefore, we argue that it is better to keep all the possible experiments to optimally cover
the input variable space (which can be highly dimensional in real problems).
In practice, quasi-Monte Carlo sequences will be preferred to pure Monte Carlo samples \citep{fanli06}.

In this section, the GLM, GAM and Gp model 
(with their relative joint extensions) are compared (see Table~\ref{tab:metamodels}). 
 To compare the predictivity of different metamodels, we use
the predictivity coefficient $Q_2$, which is the determination
coefficient $R^2$ computed from a test sample (composed here by
$10000$ randomly chosen points).  For each joint model, $Q_2$ is
computed on the mean component. 

\begin{table*}[ht]
\centering
{\scriptsize
  \caption{\small Results for the fitting of different metamodels for
    the Ishigami function. $D_{\mbox{\small expl}}$ (the explained deviance 
    of the model) and $Q_2$ (the predictivity coefficient of the model) are expressed in percent.
    For the joint models, $D_{\mbox{\small expl}}$ and $Q_2$ are those of the mean component $Y_m$.
    In the formulas for GAM, $s_1(\cdot)$, $s_2(\cdot)$ and  $s_{d1}(\cdot)$ are three spline terms.}
 \label{tab:metamodels}
  \begin{tabular}{clcclc}
  &&&&&\\
    \hline
    && $D_{\mbox{\small expl}}$ & $Q_2$ && Formula \\
    \hline
    Simple GLM && 61.3 & 60.8 && $Y = 1.92 + 2.69 X_1 + 2.17 X_2^2 -0.29 X_1^3 -0.29 X_2^4$ \\
    Joint GLM && 61.3 & 60.8 && $Y_m = 1.92 + 2.69 X_1 + 2.17 X_2^2 -0.29 X_1^3 -0.29 X_2^4$ \\
    && & && $\log(Y_d)=1.73$ \\
    &&&&&\\
    Simple GAM && 76.8 & 75.1 && $Y = 3.76 - 2.67 X_1 + s_1(X_1) + s_2(X_2)$ \\
    Joint GAM && 92.8 & 75.5 && $Y_m = 3.75 - 3.06 X_1 + s_1(X_1) + s_2(X_2)$ \\
    && & && $\log(Y_d) = 0.59 + s_{d1}(X_1)$\\
    &&&&&\\
    Simple Gp && --- & 75.0 && --- \\
    Joint Gp && --- & 75.0 && --- \\
    \hline
  \end{tabular}
}
\end{table*}

The simple GLM is a fourth order polynomial.  Only the explanatory
terms are selected in our regression model using analysis of deviance
and the Fisher statistics. The Student test on the regression
coefficients and residuals graphical analysis make it possible to
judge the goodness of fit.  We see that it remains $39\%$ of non
explained deviance due to the model inadequacy and/or to the
uncontrollable variable.  The mean component of the joint GLM gives
the same model as the simple GLM.  For the dispersion component, using
analysis of deviance techniques, no significant explanatory variable
was found. Thus, the dispersion component is supposed to be constant;
and the joint GLM is equivalent to the simple GLM approach - but with
a different fitting process.

Studying now the non parametric modeling, we start by the simple GAM
fitting where we have kept some parametric terms by applying a term
selection procedure.  The predictivity coefficient of the mean
component of the joint GAM is slightly better than the predictivity
coefficient of the simple GAM.  However, the explained deviance given
by the joint GAM mean component is clearly larger than the one given
by the simple GAM approach.  Even if this could be related to an
increasing number of parameters, as the number of parameters remains
very small compared to the data size (1000), it is certainly explained
by the fact that GAMs are more flexible than GLMs. This demonstrates
the efficiency of the joint modeling of the mean and dispersion when
heteroscedasticity is involved. Indeed, the joint procedure leads to
appropriate prior weights for the mean component. The
joint GAM improves both the joint GLM and simple GAM approaches:\\
(a) due to the GAMs flexibility, the explanatory variable $X_1$ is
identified to model the dispersion component (the interaction between
$X_1$
and the uncontrollable variable $X_3$ is therefore retrieved);\\
(b) the joint GAM explained deviance ($93\%$) for the mean component
is clearly larger than the simple GAM and joint GLM ones (joint GLM:
$61\%$, simple GAM: $77\%$).

For the Gp metamodel fitting, we use the methodology of
\citet{marioo07} which include in the model a linear
regression part and a Gp defined by a generalized exponential
covariance.
We obtain for the simple Gp 
the predictivity coefficient $Q_2=75.0\%$, which is
extremely close to the one of the simple GAM ($Q_2=75.1\%$).
The variance of the nugget effect (additional error with constant variance)
introduced in the Gp model is estimated to $25.9\%$ of the
total variance, which is close to the expected value ($1-Q_2=25.0\%$).
We can also fit, at present, a Gp model on the squared residuals to obtain a
joint Gp model (cf. section \ref{secother}).
In order to understand which inputs act in the dispersion
component, we compute the Sobol sensitivity indices of the dispersion component
 using a Monte Carlo algorithm: $S_{Y_d}(X_1)=0.996$ and
  $S_{Y_d}(X_2)=0.001$. These results draw the same conclusion
  than those obtained from the dispersion component equation of the joint GAM:
  $X_2$ is not an explanatory factor for the dispersion.
  This also leads to the right conclusion that only $X_1$ interacts with the uncontrollable 
  variable $X_3$ in the Ishigami function (\ref{eq:Ishigami}).



\begin{figure*}[ht]
  \centering
  \includegraphics[angle=-90,width=\textwidth]{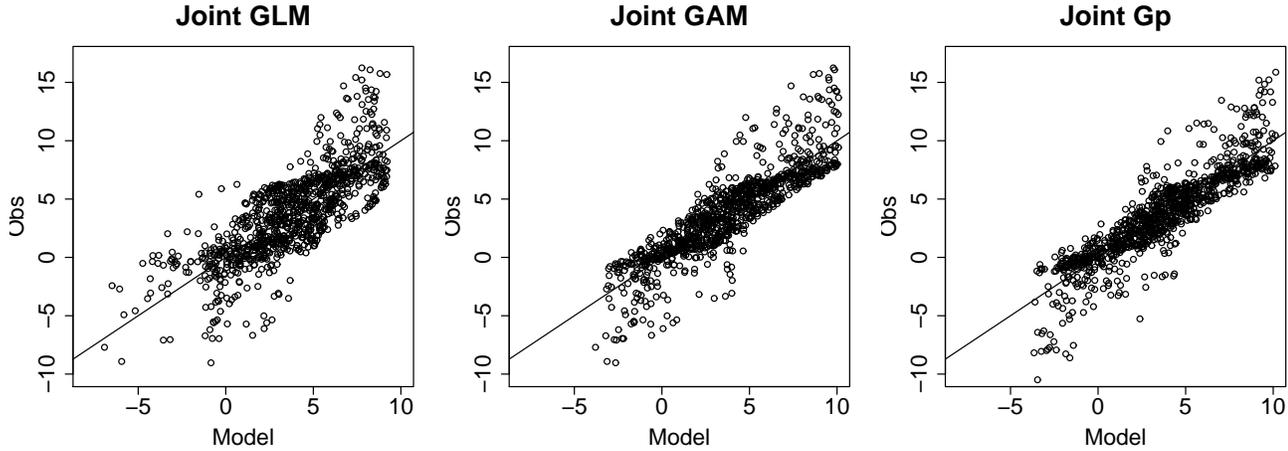}
  
  \vspace{-3cm}
  \caption{\small Observed response variable versus the predicted
    values for the three joint models: Joint GLM, Joint GAM and joint Gp
    (Ishigami application). 
  }
  \label{fig:obsVsTheoIshigami}
\end{figure*}

\begin{figure*}[ht]
  \centering
  \includegraphics[angle=-90,width=11cm]{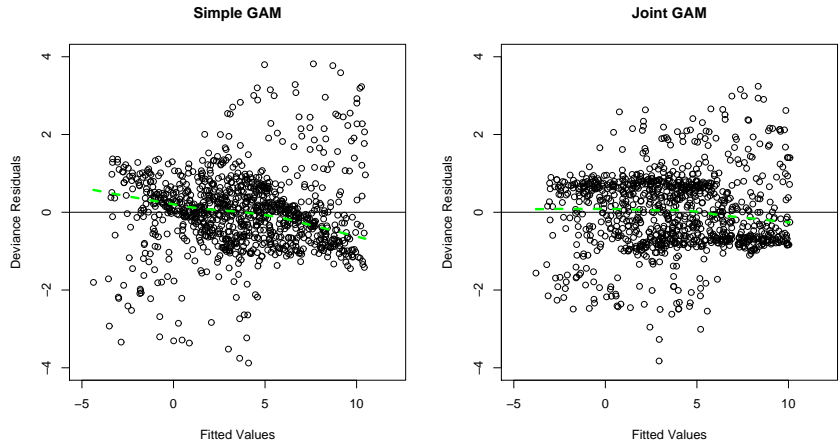}
  \caption{\small Deviance residuals for the simple and
    joint GAMs versus the fitted values (Ishigami application).
    Dashed lines correspond to local polynomial smoothers.}
  \label{fig:devResIshigami}
\end{figure*}

Let us now perform some graphical analyses in order to compare 
the results for the three joint models Joint GLM,
joint GAM and Joint Gp.
Figure~\ref{fig:obsVsTheoIshigami} shows the observed response 
against the predicted values for the three models.
First, the advantage of the GAM and Gp approaches are visible in the
Figure~\ref{fig:obsVsTheoIshigami} as the dispersion around the $y=x$
line is clearly reduced compared to the joint GLM dispersion.
Graphical comparisons between Joint GAM and Joint Gp results do not provide
any advantage for one particular model: similar biases are shown.
Second, using the GAM model, Figure~\ref{fig:devResIshigami} compares
the obtained residuals of a non parametric simple model (homoscedastic)
with the obtained residuals of a non parametric joint model (heteroscedastic).
The deviance residuals for the mean component of the joint GAM
seem to be more homogeneously dispersed around the $x$-axis; leading
to a better prediction on the whole range of the observations. 
Thus, the joint approach is more competitive than the simple one.
From this simple graphical analyses, we conclude that
a non parametric joint model (GAM or Gp) has to be preferred
to other models (simple and/or parametric).

%
%

\begin{figure*}[ht]
  \centering
  \includegraphics[angle=-90,width=16cm]{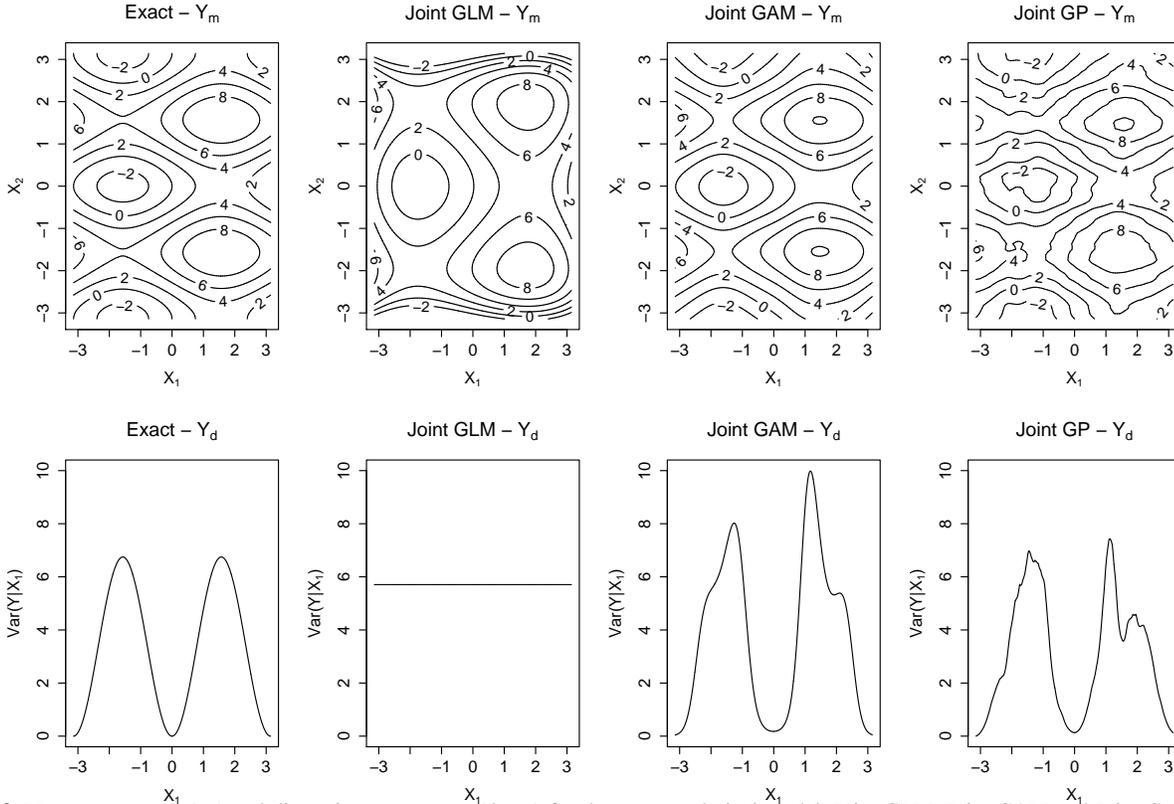}
  \vspace{-0.5cm}
  \caption{\small Mean component (up) and dispersion component (down)
    for the exact analytical model, Joint GLM, Joint GAM and Joint Gp
    (Ishigami application).}
  \label{fig:espvarcond_ish}
\end{figure*}

In order to make a finer comparison between GLM, GAM and Gp models, we examine how well they
predict the mean $Y_m(X_1,X_2)$ at inputs for which we have no data.
We can also compare the different dispersion models $Y_d(X_1)$.  The
exact analytical expressions of $Y_m$ and $Y_d$ are given in
Eq. (\ref{espvarishanal}).  Let us remark that we visualize $Y_d$
versus $X_1$ only because, for GLM and GAM dispersion models, there is
no dependence in $X_2$ and, for the Gp dispersion model, there is an
extremely small $X_2$-dependence (we then take $X_2=0$).
Figure~\ref{fig:espvarcond_ish} plots the theoretical $Y_m$  and $Y_d$ surfaces (left panels) 
and their estimates derived from the fitted joint GLM, joint GAM and 
Joint Gp models.
As shown before,
the joint GLM is irrelevant for the mean component and for the
dispersion component.  The joint GAM fully reproduces the mean
component, while joint Gp gives a rather good approximation, but with
small noise. Indeed, spline terms of GAM are perfect smoothers while
Gp predictor is impacted by residual noise on the observations: the
nugget effect does not allow to suppress all the noise induced by the
uncontrollable variable.  For the dispersion component, joint GP and
joint GAM give result of the same quality: these models correctly
reproduce the overall behaviour but with small inadequacies, probably
caused by overfitting problems.  For the two dispersion models, fitted
observations have been taken from the residuals of the mean component
learning sample.  It would be convenient, in a future work, to test
another solution by taking predicted residuals, for example by
applying a cross validation procedure.

We conclude that the joint GAM and joint Gp both adequately model the
stochastic analytical model (the Ishigami function (\ref{eq:Ishigami})).  We let some fine
comparisons between joint GAM and joint Gp for another study including
a relevant analytical application. For example, an analytical model
with strong and high order interactions will probably show the
superiority of the Gp joint model (because spline high order
interaction terms are difficult to include inside a GAM).  Therefore, in
the industrial application of section \ref{sec:appl_marthe}, we only
use the models based on GLM and GAM, while Gp could also be applied.

\subsubsection{Sobol indices}

Table~\ref{tab:ishi} depicts the Sobol sensitivity indices for the
joint GLM, the joint GAM and joint Gp using equations (\ref{eqSiY}) and
(\ref{eqSTeps}). The standard deviation estimates ($sd$) are obtained
from $100$ repetitions of the Monte-Carlo estimation procedure (which
uses $10^4$ model computations for one index estimation).  When this
Monte-Carlo procedure is used to estimate the Sobol index, we
report ``MC'' in the ``Method'' column; while ``Eq'' (resp. ``$S_{Y_d}$'')
indicates that the sensitivity indices have been deduced from the joint model
regressive equations (resp. from the sensitivity analysis of the dispersion $Y_d$).
  Therefore, no estimation errors ($sd$) are
associated to these indices (except for total indices $S_{T_i}$ which
can be deduced from $S_i$).  
When no quantitative deduction on the
sensitivity index can be made with this process, we have put a
variation interval which borders the true value. These variation
intervals are deduced from the elementary relations between 
sensitivity indices (e.g. $S_1 \leq S_{T_1}$, $S_{13} \leq S_{T_3}$, etc).

The joint GLM gives only a good estimation of $S_1$, while $S_2$ and
$S_{T_3}$ are badly estimated (errors greater than $30\%$).  $S_{12}$
is correctly estimated to zero by looking directly at the joint GLM
mean component formula (see Table \ref{tab:metamodels}).  However,
some conclusions drawn from the GLM dispersion component formula
(which is a constant) are wrong.  As no explanatory variable is
involved in this formula, the deduced interaction indices are equal to
zero: $S_{13}=S_{23}=S_{123}=0$.  Thus, $S_3=S_{T_3}=0.366$ while the
correct values of $S_3$ and $S_{T_3}$ are respectively zero and
$0.243$.

Contrary to the joint GLM, the joint GAM and joint Gp give good approximations of
all the Sobol indices. Their largest errors concern $S_{T_3}$
for the joint GAM ($7\%$-error) and joint Gp ($16\%$-error).
Moreover, the deductions drawn from the model formulas (see Table \ref{tab:metamodels})
are correct ($S_{T_2}=S_2$, $S_{12}=S_{23}=S_{123}=0$).  The only
drawback of this joint model-based method is that some indices remain unknown due to the
non separability of the dispersion component effects.  However, it can
be deduced that $S_{13}$ is non null due to the explicative effect of
$X_1$ in the dispersion component. The deduced interval variations
provide also useful information concerning the potential influence of the
interactions.

\begin{table*}[ht]
\centering
{\scriptsize
  \caption{\small Sobol sensitivity indices (with standard deviations) for
    the Ishigami function: exact and estimated values from joint GLM and
    joint GAM. ``Method'' indicates the estimation method: 
    MC for the Monte-Carlo procedure,
    Eq for a deduction from the model equations and $S_{Y_d}$ for
    a deduction from the sensitivity analysis of $Y_d(\mathbf{X})$. }
 \label{tab:ishi}
  \begin{tabular}{cclccclccclccc}
  &&&&&&&&&&&&\\
    \hline
    \multirow{2}*{Indices} & Exact &&
    \multicolumn{3}{c}{Joint GLM} && \multicolumn{3}{c}{Joint GAM} && \multicolumn{3}{c}{Joint Gp} \\
   \cline{4-6} \cline{8-10} \cline{12-14}
    & Values && Values & $sd$ & Method && Values & $sd$ & Method && Values & $sd$ & Method \\
    \hline
    $S_1$ & 0.314 && 0.314 & 4e-3 & MC && 0.325 & 5e-3 & MC  && 0.292 & 7e-3 & MC\\
    $S_2$ & 0.442 && 0.318 & 5e-3 & MC && 0.414 & 5e-3 & MC && 0.417 & 7e-3 & MC \\
    $S_{T_3}$ & 0.244 && 0.366 & 2e-3 & MC && 0.261 & 2e-3& MC && 0.205 & 1e-3 & MC \\
    $S_{12}$ & 0 && 0 & --- & Eq && 0 & --- & Eq && 0.004 & 7e-3 & MC \\
    $S_{13}$ & 0.244 && 0 & --- & Eq && $]0,0.261]$ & --- & Eq && $]0,0.205]$ & --- & $S_{Y_d}$ \\
    $S_{23}$ & 0 && 0 & --- & Eq && 0 & --- & Eq && 0 & --- & $S_{Y_d}$  \\
    $S_{123}$ & 0 && 0 & --- & Eq && 0 & --- & Eq && 0 & --- & $S_{Y_d}$ \\
    $S_{T_1}$ & 0.557 && 0.314 & 4e-3 & Eq && $]0.325,0.586]$ & ---& Eq && $]0.292,0.497]$ & ---& $S_{Y_d}$ \\
    $S_{T_2}$ & 0.443 && 0.318 & 5e-3 & Eq && 0.414 & 5e-3 & Eq && 0.417 & 7e-3 & $S_{Y_d}$ \\ 
    $S_3$ & 0 && 0.366 & 2e-3 & Eq && $[0,0.261]$ & --- & Eq && $[0,0.205]$ & --- & $S_{Y_d}$ \\
    \hline
  \end{tabular}
}
\end{table*}

Table~\ref{tab:ishi2} gives the Sobol indices computed by the same
Monte-Carlo procedure using two classical metamodels as the simple GAM
and the simple Gp. To estimate the first order Sobol indices
$S_i=V_i(Y_m)/\mbox{Var}(Y)$ (for $i=1,2$), the metamodel is used to
compute $V_i(Y_m)$ and the observed data (the $1000$ observations of $Y$)
 to compute $\mbox{Var}(Y)$.  To
estimate the total sensitivity index $S_{T_3}$ of the uncontrollable
variable, the metamodel predictivity coefficient $Q_2$ is used.
In fact, by supposing that the metamodels fit correctly the computer
code, one deduces that all the unexplained part of these metamodels is
due to the uncontrollable variable: $S_{T_3}=1-Q_2$. This is a strong
hypothesis, which is verified here due to the simplicity of the
analytical function. However, it will not be satisfied for all
application cases: in practical and complex situations, the
$Q_2$ estimation (usually done by a cross-validation method) can be
difficult and subject to caution.  For the Ishigami function, $S_1$,
$S_2$, $S_{T_3}$ are correctly estimated.  $S_{12}$ can be deduced from
the formula for the simple GAM (see Table \ref{tab:metamodels}) and estimated by
Monte-Carlo method for the Gp model.  However, any other sensitivity indices
can be proposed as no dispersion modeling is involved.

\begin{table*}[ht]
\centering
{\scriptsize
  \caption{\small Sobol sensitivity indices (with standard deviations) for
    the Ishigami function: exact and estimated values from 
    simple GAM and simple Gp model. ``Method'' indicates the estimation method: 
    MC for the Monte-Carlo procedure,
    Eq for a deduction from the model equations and $Q_2$ for the deduction
    of the predictivity coefficient $Q_2$.}
 \label{tab:ishi2}
  \begin{tabular}{cclccclccc}
  &&&&&&&&\\
    \hline
    \multirow{2}*{Indices} & Exact &&
    \multicolumn{3}{c}{Simple GAM} && \multicolumn{3}{c}{Simple Gp} \\
    \cline{4-6} \cline{8-10} 
    & Values && Values & $sd$ & Method && Values & $sd$ & Method \\
    \hline
    $S_1$ & 0.314 && 0.333 & 6e-3 & MC && 0.292 & 7e-3 & MC \\
    $S_2$ & 0.442 && 0.441 & 6e-3 & MC && 0.417 & 7e-3 & MC \\
    $S_{T_3}$ & 0.244 && 0.249 & --- & $Q_2$ && 0.250 & --- & $Q_2$ \\
    $S_{12}$ & 0 && 0 & --- & Eq && 0.004 & 7e-3 & MC \\
    \hline
  \end{tabular}
}
\end{table*}

{\it 
  \noindent{\bf \itshape Remark:}
  Estimating the nugget effect variance of the Gp model mean component
  gives another estimation of the total sensitivity index of the
  uncotrollable variable.  In this example, the variance of the
  nugget effect has been estimated to $25.9\%$ of the total variance,
  which is close to the exact value ($24.4\%$).  However, this
  estimation can be difficult in more complex situations, because of a
  difficult optimization step while fitting the Gp model \citep{fanli06,marioo07}.
}

In conclusion, this example shows that the joint non parametric models
 can adjust complex heteroscedastic situations for which
classical metamodels are inadequate.  Moreover, the joint models offer
a theoretical basis to compute efficiently global sensitivity indices of
stochastic models.


\subsubsection{Convergence study}
 
\begin{figure*}[ht]
  \centering
  \includegraphics[angle=-90,width=14cm]{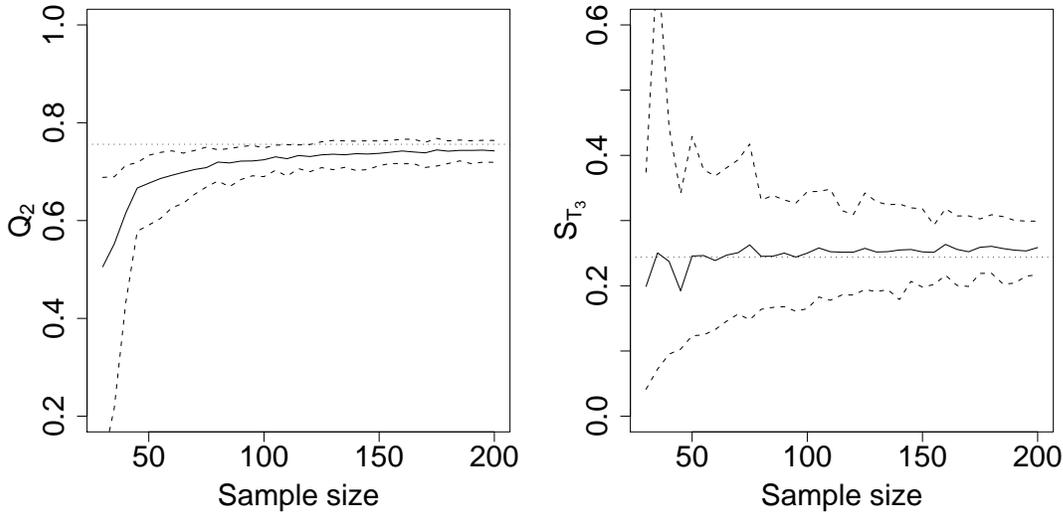}
  \caption{\small For the Ishigami function, mean and
    $90\%$-confidence interval (based on $100$ replicates) of joint
    GAM $Q_2$ and $S_{T_3}$ in function of the learning sample size
    $n$.}
  \label{fig:cvence}
\end{figure*}

In order to provide some practical guidance for the sampling size
issue, we perform a convergence study for the estimation of the joint
GAM and the associated sensitivity indices.  Figure~\ref{fig:cvence}
shows some convergence results for a learning sample size $n$ varying
between $30$ to $200$ by step of $5$.
The learning points are sampled by the simple Monte Carlo technique.
  The predictivity coefficient
$Q_2$ is obtained from a test sample (composed of $1000$ randomly
chosen points).  The total sensitivity index of the uncontrollable
variable $S_{T_3}$ is obtained by averaging the dispersion component
$Y_d$ (with $1e6$ randomly chosen points).  We can notice the rapid
convergence of the predictivity coefficient $Q_2$ and the slower
convergence of $\mathbb{E}(Y_d)$.  The convergence speed for $S_1$ and
$S_2$ computed from the mean component are not shown here but are
similar to the one of $Q_2$.

From this particular case (low-dimensional but rather complex
numerical model due to non linearities and strong interaction), we
conclude that a $100$-size sample is sufficient for fitting the joint
GAM and for obtaining precise sensitivity indices.  Moreover, for the
estimation of the total sensitivity index of the uncontrollable
variable, using the predictivity coefficient of the mean component is
highly recommended (instead of using the dispersion component).  With
additional experiments, \citet{ioorib08} have
confirmed this result.
 
In practice, the way to ensure that the convergence has been reached would be to
visualize $Q_2$ and its confidence interval (by a bootstrap technique for example)
 by resampling in the learning sample and progressively increasing its size.

\subsection{Application to an hydrogeologic transport code}\label{sec:appl_marthe}

The joint approach is now applied to a complex industrial model of
radioactive pollutants transport in saturated porous media using the
MARTHE computer code (developed by BRGM, France).
In the context of an environmental impact study, MARTHE has been
applied to a model of strontium 90 ($^{90}$Sr) transport in saturated
media for a radwaste temporary storage in Russia \citep{volioo08}.
Only a partial characterization of the site has been made and,
consequently, values of the model input variables are not known
precisely: 20 scalar input variables have been considered as random
variables, each of them associated to a specified probability density
function.  The model output variables of interest concern the
$^{90}$Sr concentration values in different spatial locations. One of
the main goals of this study is to identify the most influential
variables of the computer code in order to improve the
characterization of the site in a judicious way.  Because of large
computing times of the MARTHE code, the Sobol sensitivity indices are
computed using metamodels (\citealp[boosting regression trees model for][]{volioo08}
\citealp[and Gaussian process model for][]{marioo07}).

As a perspective of the \citet{volioo08} work, 
\citet{ioo08} studies more precisely the influence of the spatial form of an
hydrogeologic layer. The method consists in performing a geostatistical
simulation of this layer (which is a two-dimensional spatial random
field), before each calculation of the computer model.  This
geostatistical simulation is rather complex and the resulting spatial
field cannot be summarized by a few scalar values.  Therefore, as
explained in our introduction, this hydrogeologic layer form has to be
considered as an uncontrollable variable of the computer model.
Additionally to the uncontrollable variable, $16$ scalar input
variables remain uncertain and are treated as random variables.  It
concerns the permeability of different geological layers, the
longitudinal and transversal dispersivity coefficients, the sorption
coefficients, the porosity and meteoric water infiltration
intensities.

In order to keep coherence with \citet{volioo08} previous study,
the learning sample size has been chosen to be the same: $N=300$.
This size is in adequation with the heuristic recommandation of $10$ observations
per input dimension \citep{loesac08,marioo07}, used in most of the practical studies 
on deterministic computer codes.
The Latin Hypercube Sampling method is used to obtained a sample
of $N$ random vectors (each one of dimension $16$). In addition, $N$
independent realizations of the spatial random field (noticed by
$\varepsilon$) are obtained by a specific geostatistical simulation
algorithm \citep{ioo08}.  
Performing independent realizations for each of the
simulator run has been imposed by the small number of available runs ($300$)
relatively to the high-dimensional model ($20$). Moreover, one
of our primary concern was also to perform an uncertainty propagation study,
in which replicates have to be avoided.
In any case, more interesting designs should be chosen, making
replicates for example by changing
the controllable input factors while keeping fixed the geostatistical realization.
However, such ideas are well beyond the scope of the current paper
\citep[see][for a recent review about the design issue]{andbor09}.

After $8$ calculation days, we obtain $300$ observations 
of the output variable of the MARTHE model
 ($^{90}$Sr concentration at the domain center).
As two computer runs have given incoherent values, 
we keep $298$ observations. For the GLMs and
GAMs construction phase, the large data dispersion suggests the use of
logarithmic link functions for $g$ and $h$ (see Eqs (\ref{eqGlmM})
and (\ref{eqGlmD})).  Due to the large number of inputs, a manual term
selection process has been applied.  No interaction term has been
found to be explicative in the GLMs.  However, a bi-dimensional spline
term has been added in the GAMs because of convincing deviance
contribution and negligible p-value.
To find this significant interaction term, we have not introduced 
in the model all the $120$ interaction terms. We have sequentially 
tested all the interaction terms involving one significant 
first order term ($kd1$, $kd2$, $per2$ and $per3$) and each other factor.
Then, we keep the interaction terms which show some explanatory contribution
to the model.
  
The results are summarized below by
giving the explained deviance and the explanatory terms involved in
the formulas:
\begin{itemize}
\item Simple GLM: $D_{\mbox{\small expl}}=60\%$ with the terms $kd1$,
  $kd2$, $per1$, $per2$.
\item Joint GLM: $D_{\mbox{\small expl}}(\mbox{mean})=66.4\%$, with
  the same terms than the simple GLM, $D_{\mbox{\small
      expl}}(\mbox{dispersion})=8.7\%$ with the terms $kd1$ and
  $per3$.
\item Simple GAM: $D_{\mbox{\small expl}}=81.8\%$ with $s(kd1)$,
  $s(kd2)$, $s(per3)$, $s(per2,kd2)$.
\item Joint GAM: $D_{\mbox{\small expl}}(\mbox{mean})=98.1\%$ with the
  same terms than the simple GAM, $D_{\mbox{\small
      expl}}(\mbox{dispersion})=29.7\%$ with $kd1$, $kd2$.
\end{itemize}
$kd1$, $kd2$ and $per1$, $per2$, $per3$ are respectively the sorption
coefficients and the permeabilities of the different hydrogeologic layers.
One observes that the GAM models outperform the GLM ones. 
The predictivity coefficient (computed
by the leave-one-out method) of the simple GAM gives $Q_2=76.4\%$,
while for the simple GLM $Q_2=58.8\%$.

\begin{figure*}[ht]
  \centering
  \includegraphics[angle=-90,width=\textwidth]{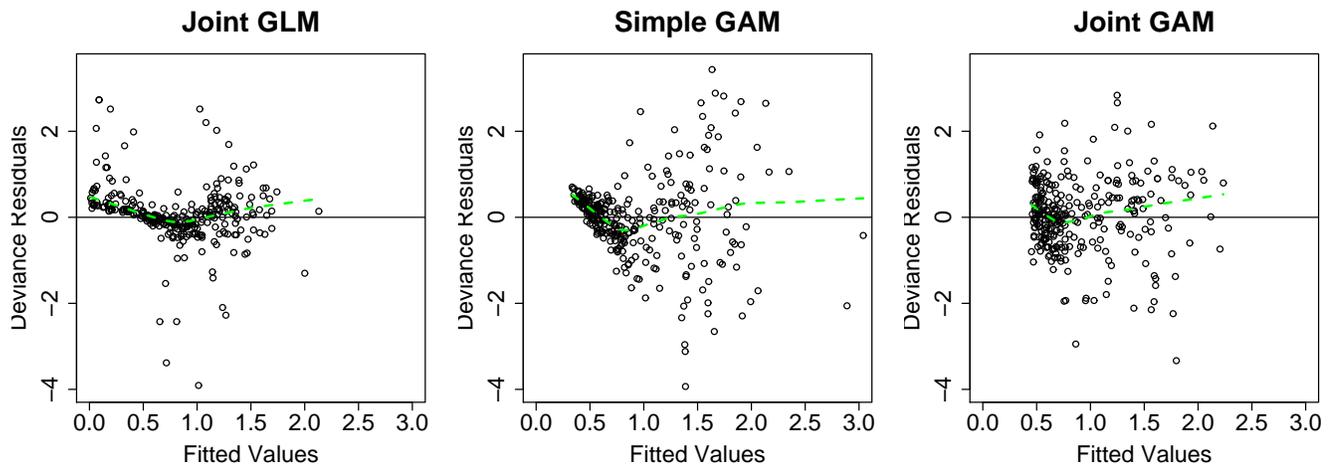}
  \caption{\small Deviance residuals (mean component) for the Simple GAM,
    Joint GAM and Joint GLM versus the fitted values (MARTHE
    application). Dashed lines correspond to local polynomial
    smoothers.}
  \label{fig:devResMarthe}
\end{figure*}

Figure~\ref{fig:devResMarthe} shows the deviance residuals
against the fitted values for the joint GLM, simple GAM and joint GAM
models. 
For the joint GLM approach, some outliers are not visible to
keep the figure readable. As a consequence, the GAMs clearly
lead to smaller residuals.
Moreover, the joint GAM outperforms the simple GAM due to the right
explanation of the dispersion component. It can be seen that the joint
GAM allows to suppress the bias involved by the heteroscedasticity, while
simple GAM residuals are affected by this bias.

\begin{figure*}[ht]
  \centering
  \includegraphics[angle=-90,width=\textwidth]{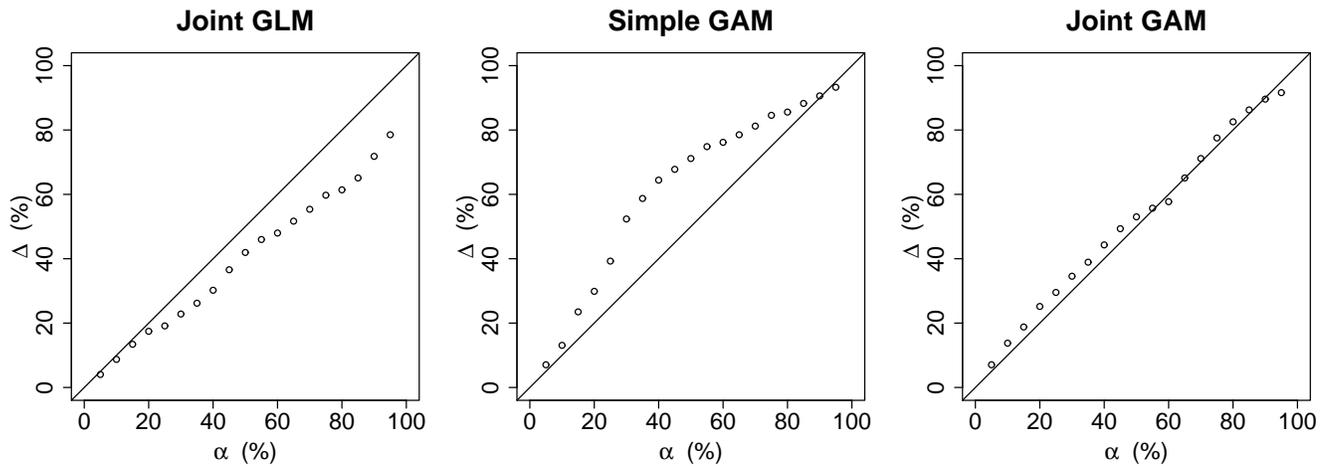}
  \caption{\small Proportion $\Delta$ (in percent) of observation that lie
    within the $\alpha$\% theoretical confidence interval in function
    of the confidence level $\alpha$. MARTHE application.}
  \label{fig:deltaVsAlphaMarthe}
\end{figure*}

Figure~\ref{fig:deltaVsAlphaMarthe} shows the proportion $\Delta$ of
observations that lie within the $\alpha$\% theoretical confidence
interval against the confidence interval $\alpha$. By definition,
if a model is suited for both mean and dispersion modelings, the
points should be located around the $y=x$ line. As a consequence, this
plot is useful to compare the goodness of fit for the different
models. It can be
seen that the joint GAM is clearly the most accurate model. Indeed, 
all its points are close to the theoretical $y=x$ line, while the joint
GLM (resp. simple GAM) systematically leads to underestimations (resp.
overestimations).
Consequently, from the
Figures~\ref{fig:devResMarthe}--\ref{fig:deltaVsAlphaMarthe}, one deduces that the
joint GAM model is the most competitive one. On one hand, the mean
component is modeled accurately without any bias. On the other hand,
the dispersion component is competitively modeled leading to reliable
confidence intervals.


Table~\ref{tab:marthe} gives the main Sobol sensitivity indices for
the joint GLM, joint GAM and simple GAM (using $10^4$ model
computations for one index estimation). The Sobol indices of the
interactions between controllable variables are not given (except
between $kd2$ and $per2$) because these interactions are not included
in the formulas of the two joint models. Therefore, their Sobol
indices are zero. The two joint models give similar results for all
first order sensitivity indices. The sorption coefficient of the
second layer $kd2$ explained more than $52\%$ of the output variance,
while the permeability of the second layer $per2$ explained more than
$5\%$.  Some large differences arise in the total influence of the
uncontrollable variable $\varepsilon$: $38.2\%$ for the joint GLM and
$27.7\%$ for the joint GAM.  Moreover, the joint GLM shows an
influence of the interaction between $per3$ and $\varepsilon$, while
the joint GAM shows an influence of the interaction between $kd2$ and
$\varepsilon$.  In this application, we consider the joint GAM results
more reliable than the joint GLM ones because the joint GAM captures
more efficiently the mean and dispersion components of the data than
the joint GLM.

\begin{table*}[ht]
{\scriptsize
  \centering
  \caption{\small Estimated Sobol sensitivity indices (with standard
    deviations obtained by $100$ repetitions) for the MARTHE
    code. ``Method'' indicates the estimation method: 
    MC for the Monte-Carlo procedure,
    Eq for a deduction from the model equations and $Q_2$ for the deduction
    of the predictivity coefficient $Q_2$. ``---'' indicates that the value is not available.}
    
 \vspace{0.2cm}
  \label{tab:marthe}
  \begin{tabular}{llccclccclccc}
    \hline
    \multirow{2}*{Indices} &&
    \multicolumn{3}{c}{Joint GLM} && \multicolumn{3}{c}{Joint GAM}&& \multicolumn{3}{c}{Simple GAM} \\
    \cline{3-5} \cline{7-9} \cline{11-13}
    && Values & $sd$ & Method && Values & $sd$ & Method && Values & $sd$ & Method \\
    \hline
    $S$(kd1)  && 0.002 & 0.6e-2 & MC && 0.037 & 1.0e-2 & MC && 0.140 & 1.0e-2 & MC \\
    $S$(kd2) && 0.522 & 0.6e-2 & MC && 0.524 & 1.0e-2 & MC && 0.550 & 1.1e-2 & MC \\
    $S$(per1) && 0.018 & 0.7e-2 & MC && 0 & --- & Eq && 0 & --- & Eq \\
    $S$(per2) && 0.052 & 0.6e-2 & MC && 0.078 & 1.0e-2 & MC  && 0.044 & 1.0e-2 & MC \\
    $S$(per3) && 0 & --- & Eq && 0.005 & 1.0e-2 & MC && 0.008 & 1.0e-2 & MC \\
    $S$(kd2,per2)  && 0 & --- & Eq && 0.063 & 1.0e-2 & MC && 0.026 & 1.0e-2 & MC \\
    $S_{T}(\varepsilon)$ && 0.382 & 0.2e-2 & MC && 0.277 & 0.3e-2 & MC && 0.235 & --- & $Q_2$ \\
    $S$(kd1,$\varepsilon$) && $]0,0.382]$ & --- & Eq && $]0,0.277]$ & --- & Eq && --- & --- & --- \\
    $S$(kd2,$\varepsilon$) && 0 & --- & Eq && $]0,0.277]$ & --- & Eq && --- & --- & --- \\
    $S$(per1,$\varepsilon$) && 0 & --- & Eq && 0 & --- & Eq && --- & --- & --- \\
    $S$(per2,$\varepsilon$) && 0 & --- & Eq && 0 & --- & Eq && --- & --- & --- \\
    $S$(per3,$\varepsilon$) && $]0,0.382]$ & --- & Eq && 0 & --- & Eq && --- & --- & --- \\
    \hline
  \end{tabular}
  }
\end{table*}

By comparing the joint GAM results with the simple GAM
results, some significant differences can be printed out:
\begin{itemize}
\item The $kd1$ first order sensitivity index is overestimated using
  the simple GAM ($14.0\%$ instead of
  $3.7\%$ for the joint GAM).  Indeed, the deviance analysis of the
  joint GAM dispersion component shows a high contribution of $kd1$,
  which means that the interaction between $kd1$ and the
  uncontrollable variable is probably large.  For a standard
  metamodel, like the simple GAM, this interaction is not
  found out and leads to a wrong estimation of the first order
  sensitivity index of $kd1$.
\item For the simple metamodels, using the relation
  $S_{T}(\varepsilon)=1-Q_2$, the total sensitivity index of the
  uncontrollable variable is underestimated: $23.5\%$ (simple GAM)
  instead of $27.7\%$ (joint GAM).  The
  classical metamodels tend to explain some parts of the data which
  can be adequately included in the dispersion component of the joint
  GAM during the iterative fitting algorithm.
\item Contrary to the other metamodels, the joint GAM allows to prove
  that only $kd1$ and $kd2$ interact with the uncontrollable variable.
\end{itemize}

As a conclusion, these sensitivity analysis results will be very useful to the
physicist or the modeling engineer during the model construction and
calibration steps.  In this specific application, the sensitivity
analysis shows that the geometry of the second hydrogeological
layer has a strong influence (up to $28\%$) on the predicted $^{90}$Sr
concentration.  Therefore, an accurate modeling
of this geometry, coupled with a better knowledge of the most
influential variable $kd2$, are the key steps to an important
reduction of the model prediction uncertainties.

%


\section{Conclusion}

This paper has proposed a solution to compute variance-based
sensitivity indices of stochastic computer model outputs. It consists
in modeling the mean and the dispersion of the code outputs by two
explanatory models.  The classical way is to separately build these
models.  In this paper, the use of the joint modeling is preferred.
This theory, proposed by \citet{pre84} and \citet{smy89},
then extensively
developed by \citet{nel98a}, is a powerful tool to fit the mean
and dispersion components simultaneously. \citet{zabdej98} already applied this approach to model stochastic
computer code.  However, the behavior of some numerical models can be
highly complex and non linear.  In the present paper, some examples
show the limit of this parametric joint model.  Being inspired by
\citet{rigsta96} who use non parametric joint
additive models (restricted to Gaussian cases), we have developed a
more general joint model using GAMs and quasi distributions. Like
GLMs, GAMs are a suited framework because it allows variable and model
selections {\it via} quasi-likelihood function, classical statistical
tests on coefficients and graphical displays.  Additional works using
joint GLMs and joint GAMs for computer experiments can be found in
\citet{ioorib08}.

The joint GAM has proven its flexibility to fit complex data: 
we have obtained the same performance for its mean and dispersion
components as the powerful Gp model.
Dealing with computer codes involving many factors and
strong interactions between model factors, 
it would be convenient to look more precisely at other joint models, as
the joint Gp model we have shortly described and used.
An analytic case on the Ishigami function shows that these two 
non parametric joint models (GAM and Gp) are adapted to
complex heteroscedastic situations where classical metamodels
 are inadequate.  Moreover, it offers a theoretical basis to
compute Sobol sensitivity indices in an efficient way. 
The analytical formulas available with the joint GAM are very useful
to complete the sensitivity analysis results and to improve our model
understanding and knowledge.

The performance of the joint model approach was assessed on an
industrial application.  Compared to other methods, the
modeling of the dispersion component allows to obtain a robust
estimation of the total sensitivity index of the uncontrollable
variable, which leads to correct estimations of the first order
indices of the controllable variables.  In addition, it reveals the
influential interactions between the uncontrollable variable and the
other input variables. Obtaining quantitative values for these
interaction effects is still an open issue, but a challenging problem. 
Finally, the joint model would also serve in the uncertainty
propagation studies of complex models, to obtain the full distribution
of the model output.

In the whole, all statistical analysis were performed using the R
software environment \citep{Rsoft}. In particular, the following
functions and packages were useful: the ``glm'' function to fit a
simple GLM, the ``mgcv'' (Multiple Smoothing Parameter Estimation
by GCV) package to fit a simple GAM, and the ``sensitivity'' package
to compute Sobol indices. We also developed the
``JointModeling'' package to fit joint models (including joint GLM and
joint GAM).

\bibliographystyle{apalike}

\end{document}